\documentclass[aps,prd,superscriptaddress,10pt,noshowpacs, nofootinbib]{revtex4}

\usepackage{amsmath,amsthm,amsfonts,amssymb}
\usepackage[mathcal]{eucal}
\usepackage{mathrsfs}
\usepackage{latexsym}
\usepackage{graphicx}
\usepackage[utf8]{inputenc}
\usepackage{graphicx,epstopdf}
\usepackage{latexsym}
\usepackage{times,txfonts}
\usepackage{float}%para fixar as figuras na posição desejada

\newcommand{\sgn}{\mbox{sgn}}
\newcommand{\sech}{\;\mathrm{sech}}
\newcommand{\mc}{\mathcal}

\newcommand{\p}{\partial}

\begin{document}
\title{A Gravity-Consistent Confinement of Fermions in Braneworld}

\author{L. F. F. Freitas}
\email{luizfreitas@fisica.ufc.br} \affiliation{Departamento de F\'isica,
  Universidade Federal do Cear\'a, Caixa Postal 6030, Campus do Pici,
  60455-760 Fortaleza, Cear\'a, Brazil}

\author{I. C. Jardim}
\email{ivan.jardim@urca.br} \affiliation{Departamento de F\'{i}sica, Universidade Regional do Cariri, 57072-270, Juazeiro do Norte, Cear\'{a}, Brazil}

\author{G. Alencar}
\email{geova@fisica.ufc.br} \affiliation{Departamento de F\'isica,
 Universidade Federal do Cear\'a, Caixa Postal 6030, Campus do Pici,
 60455-760 Fortaleza, Cear\'a, Brazil}

 \author{R. R. Landim}
\email{renan@fisica.ufc.br} \affiliation{Departamento de F\'isica,
  Universidade Federal do Cear\'a, Caixa Postal 6030, Campus do Pici,
  60455-760 Fortaleza, Cear\'a, Brazil}

\begin{abstract}
In this manuscript, we discuss the confinement of the spin $\frac{1}{2}$ field on a plethora of branewords models. Recently, in (Eur.Phys.J.C 80 (2020) 5, 432), we studied the consistency of the Standard Model (SM) fields localization on braneworlds with the Einstein equation. In that paper, we discussed the consistency of the spinor field confinement and, by using a Yukawa-like interaction given by $\mc{L}_{int}\propto f(y)\bar{\Psi}\Psi$, we obtained that the function must be defined as $f(y)\propto e^{-A}A'$. This shape of the scalar function emerge from the requirement that the spin $\frac{1}{2}$ (zero-mode) localization cannot modify the metric on bulk. This ensures that the confinement of gravity on the brane is preserved. In the present manuscript, we find a covariant scalar-coupling function that can generate this interaction. This provide a new mechanism for localizing fermion fields over the brane. We also discuss massive modes and we found some gravitational configuration where there are confined and discretized massive modes.
\end{abstract}

\maketitle
%\tableofcontents
\section{Introduction}\label{Sec-1}

Braneworld models play an important role in the context of extra dimensions. Since the first papers presented by L. Randall and R. Sundrum (RS) \cite{Randall:1999vf, Randall:1999ee}, a wide variety of other brane models were proposed in different gravitational scenarios \cite{Bazeia:2003qt, Liu:2009dwa, Gregory:1999gv, Flachi:2009uq, Kehagias:2000au, Bazeia:2004dh, Bazeia:2003aw, Bazeia:2009sr, Bazeia:2020qxr, Guo:2011qt, Gherghetta:2000qt, Oda:2000zc, Giovannini:2001hh, Gogberashvili:1998iu, deCarlos:2003nq, Silva:2012yj, Arkani-Hamed:1998jmv}. For all these models, the gravitational field (zero-mode) can be localized on a $4$-dimensional brane and this allows us to recover the well-known Newton's gravitational law. Beyond the gravitational one, confinement of the Standard Model (SM) fields also play an important role in this context.

The RS models consider the Standard Model (SM) fields previously confined on a delta-like $3$-brane. However, other studies showed that this cannot be ensured for all SM fields \cite{Chang:1999nh, Bajc:1999mh, Pomarol:1999ad, Davoudiasl:1999tf}. In fact, for the free case, only the scalar field can be confined simultaneously with the gravitational field for most of those braneworlds mentioned above. The $U(1)$ gauge field cannot be localized for most the $5$ dimensional brane models \cite{Liu:2010pj, Bajc:1999mh, Pomarol:1999ad}. On the other hand, for $6$D braneworlds, gauge field can be confined for string-like or vortex brane models \cite{Midodashvili:2003ib, Oda:2000zc, Costa:2015dva, Giovannini:2002sb, Torrealba:2010sg, Costa:2013eua}, but this field still have confinement problems for models with the two extra dimensions infinitely large \cite{Randjbar-Daemi:2000lem}. Other important SM field in this context is the spinor field. For most of the braneworld models investigated up to now, the free spinor field cannot be localized \cite{Randjbar-Daemi:2000lem, Ringeval:2001cq, Melfo:2006hh}. Due to this, some localization mechanisms for this field were proposed. For example, in $5$-dimensional braneworlds, it is commonly used a Yukawa-like interaction given by $\mc{L}_{int}=-\lambda f(y)\bar{\Psi}\Psi$ \cite{Randjbar-Daemi:2000lem, Kehagias:2000au, Barbosa-Cendejas:2015qaa, Koley:2004at,Castro:2010au, Mendes:2017hmv,Zhao:2010mk,Liu:2008pi, Liu:2009ve,Liu:2009uca,Castro:2011pp, Cruz:2011ru}. Up to now, the scalar-coupling function $f(y)$ can be arbitrarily chosen since it provides the zero-mode confinement. Still in this $5$D scenarios, others aspects of the spinor field dynamics on the bulk also were discussed in the literature, for example, by looking for resonant massive modes \cite{Landim:2011ki, Landim:2011ts}.

Generally, the confinement study of the massive modes for all these fields is neglected. Nevertheless, for the spinor field there are some interesting results about this point in the literature. In references  \cite{Dubovsky:2000am, Ringeval:2001cq, Melfo:2006hh}, for example, the authors discussed the confinement in $5$D brane model by using $f(y)\propto\phi$ (kink-like scalar field). With this, they obtained a Schrödinger-like equation (asymptotic) where the `potential' has a local minimum on the brane. This allows the confinement of lighter massive modes, however this localization is not stable and these modes can tunneling to the bulk. In reference \cite{Landim:2015paa}, the authors propose a particular scenarios, where the scalar-coupling function $f(y)$ is properly chose, and they get some confined massive modes. For this case, it is obtained an analytical solution and its confinement is stable. A similar result is also obtained in reference \cite{Castro:2010uj}. However, for this last case, the brane is generated by a scalar field with a non-standard (non-quadratic) kinetic term.

Recently, in reference \cite{Freitas:2020mxr}, we discussed the consistency of the Standard Model fields localization procedure with the Einstein equation. With this, we obtained a very important constraint on the shape of the coupling function $f(y)$. Namely, it must be given by $f(y)=\lambda e^{-A}A'$
This function emerge from the consistency analysis with the Einstein equation for the confined zero-mode of the spinor field. However, this term seems to be not covariant.

In the present manuscript, we apply the procedure shown in Ref.  \cite{Freitas:2020mxr} to produce a gravity-consistent localization of the zero-mode of the fermion field. This mechanism constraints the Yukawa-coupling function and the coupling constant in terms of the metric of the brane model used. Since this field has been studied in the literature from some different brane models, we will explicitly apply the gravity-consistent condition for these models and compare them with the literature results. In addition, we intend to propose a covariant origin for this coupling in a brane driven by scalar fields, single or multiple fields, using the superpotential method. Since the constraints are obtained to provide a gravity-consistent localization of the zero-mode of the spinorial field, we will study the consequences of this result on the confinement of the massive modes for a particular $5$D thick brane models presented by \cite{Gremm:1999pj}.

This work is organized as follow. In section \ref{Main},  we present a general discussion about the spinor field localization for an arbitrary $5$D brane. And, still in section \ref{Main}, we discuss the confinement of the zero-mode in a consistent way with the Einstein's equation. Next, we apply the result obtained in previous section to the Randall-Sundrum scenario, in section \ref{Sec-3}, and for the picewise thick brane model, in section \ref{Sec-4}, in order to compare with the literature. The thick brane models driven by a single scalar field are addressed in section \ref{Sec-5} and, for each specific case, in its subsections. The section \ref{Sec-6} is reserved for a review of thick brane models driven by multiple scalar fields, and, using the superpotential method, to find a covariant way to write the scalar-coupling function which localize the zero-mode. Also, in section \ref{Sec-6}, we apply the result for some specific brane models. Finally, we study the localization of the massive modes in the braneworld \cite{Gremm:1999pj} in section \ref{Sec-7} and the conclusions are presented in section \ref{Sec-8}.

\section{A Gravity-Consistent Confinement of Fermions in Braneworld}\label{Main}

In this section, let us perform a general discussion about the confinement of the spinorial field on braneworlds. In doing this, we will consider only $5$D thick brane models, thus, let us use the warped metric
\begin{eqnarray}\label{EQ018}
ds^{2}=g_{MN}dx^{M}dx^{N}=e^{2A(y)}\hat{g}_{\mu\nu}(x)dx^{\mu}dx^{\nu}+e^{2B(y)}dy^{2},
\end{eqnarray}
where $A(y)$ and $B(y)$ are a generic warp factors. In this section, we will not consider the form of the factors. Thus, eventually, the below results can be applied for every 5-dimensional braneworlds with a generic metric (\ref{EQ018}).

To discuss the localization of the spinorial field in the above background, let us start writing the action for this field as
\begin{eqnarray}\label{EQ019}
S^{(m)}=-\int d^{4}xdy\sqrt{-g}\Bar{\Psi}\left[i\Gamma^{M}D_{M}+\lambda f(y)\right]\Psi,
\end{eqnarray}
where, $D_{M}=\p_{M}-\omega_{M}$ is the covariant derivative and $\omega_{M}=\frac{1}{4}\omega^{ab}_{M}\gamma_{a}\gamma_{b}$ are the spin connections. For a Gamma matrix $\Gamma_{M}$ ($\gamma_{a}$), index $M,N,$ etc. ($a,b,$ etc.), are defined in the curved (flat) spacetime \footnote{Throughout the manuscript, capital indexes $M,N,..$ ($a,b,..$) run on all $5$ dimensions for curved (flat) spacetime. Greek index $\mu,\nu,..$ run on the brane dimension $\mu,\nu=(1,2,3,4)$.}. Beyond this, $f(y)$ is the scalar-coupling function and it will be properly chosen later to provide the confinement of the zero-mode. With this action, we can discuss both the free case ($\lambda=0$) and also some localization mechanisms presented in the literature.

From the action (\ref{EQ019}), we can perform the variation with respect to the field $\Bar{\Psi}$ and obtain the equation of motion (EOM)
\begin{eqnarray}\label{EQ020}
\left[i\Gamma^{M}D_{M}+\lambda f(y)\right]\Psi=\left[i\Gamma^{\mu}D_{\mu}+i\Gamma^{y}D_{y}+\lambda f(y)\right]\Psi=0.
\end{eqnarray}
Gamma matrix in the curved spacetime $\Gamma^{M}$ are related to that in the flat (Minkowski) spacetime by $\Gamma^{M}(x,y)=E^{M}_{a}(x,y)\gamma^{a}$, where $E^{M}_{a}(x,y)$ are the {\it vielbein}. The vielbein must satisfy the relation $E^{M}_{a}E^{N a}=g^{MN}$ and the Gamma matrix $\Gamma^{M}$ must satisfy the Clifford algebra
\begin{eqnarray}\label{EQ021}
\Gamma^{M}\Gamma^{N}+\Gamma^{N}\Gamma^{M}=\left\lbrace\Gamma^{M},\Gamma^{N}\right\rbrace =-2g^{MN}.
\end{eqnarray}
With all this, we will choose the vielbein as
\begin{eqnarray}
\left. \begin{array}{c}
E^{\mu}_{a}(x,y)=e^{-A(y)}\hat{e}^{\mu}_{a}(x) \\
E^{y}_{a}(x,y)=0\hspace{1.8cm}
\end{array} \right\rbrace\ \ \mbox{for}\ \ a=1,2,3,4,\label{EQ022}\\
\left. \begin{array}{c}
E^{\mu}_{5}(x,y)=0\hspace{0.88cm}\\
E^{y}_{5}(x,y)=e^{-B(y)}
\end{array}\hspace{0.95cm} \right\rbrace\ \ \mbox{for}\ \ a=5.\hspace{1.12cm}\label{EQ023}
\end{eqnarray}
In these definitions, $\hat{e}^{\mu}_{a}(x)$ are the vielbein on the brane and it must satisfy $\hat{e}^{\mu}_{a}(x)\hat{e}^{\nu a}(x)=\hat{g}^{\mu\nu}(x)$. Note that, as the spinor field can be confined, the brane metric must change from $\eta_{\mu\nu}\to \hat{g}_{\mu\nu}(x)$. Afterwards, in this section , let us discuss the consistency of the localization procedure and the change in the brane metric should be clearer.

By using these vielbein, we can show that the spin connection are given by
\begin{eqnarray}\label{EQ024}
\omega_{\mu}(x,y)=\hat{\omega}_{\mu}(x)+\frac{1}{2}\Gamma_{\mu}\Gamma^{y}A',\ \ \ \omega_{y}(x,y)=0.
\end{eqnarray}
Here, {\it prime} ( $'$ ) means derivative with respect to the extra dimension $y$. Finally, the EOM can be writing as
\begin{eqnarray}\label{EQ025}
i\hat{\Gamma}^{\mu}(x)\hat{D}_{\mu}\Psi+2ie^{A-B}A'\gamma^{5}\Psi+ie^{A-B}\gamma^{5}\Psi'+\lambda e^{A}f(y)\Psi=0,
\end{eqnarray}
where $\hat{D}_{\mu}=\p_{\mu}-\hat{\omega}_{\mu}(x)$ and $\hat{\Gamma}^{\mu}(x)=\hat{e}^{\mu}_{a}(x)\gamma^{a}$ $[a=1,2,3,4]$. Now, to perform the separation of the variables, let us propose the ansatz
\begin{eqnarray}\label{EQ026}
\Psi(x,y)=\sum_{n}\Psi_{n}(x,y)=e^{-2A(y)}\sum_{n}\left[\psi^{+}_{(n)}(x)\xi^{+}_{n}(y)+\psi^{-}_{(n)}(x)\xi^{-}_{n}(y)\right],
\end{eqnarray}
where the spinorial characteristic is in quantities $\psi^{\pm}_{(n)}(x)$ and they must satisfy $-i\gamma^{5}\psi^{\pm}_{(n)}=\pm\psi^{\pm}_{(n)}$. With this, the variables can be separated as
\begin{eqnarray}
i\hat{\Gamma}^{\mu}\hat{D}_{\mu}\psi^{\pm}_{(n)}+m_{n}\psi^{\mp}_{(n)}=0,\label{EQ027}\\
-e^{A-B}\frac{d\xi^{\pm}_{n}(y)}{dy}\pm \lambda e^{A}f(y)\xi^{\pm}_{n}(y)=\pm m_{n}\xi^{\mp}_{n}(y).\label{EQ028}
\end{eqnarray}
The eqs. (\ref{EQ027}) are the equations of motion of four dimensional fermions on the brane, $\psi^{\mp}_{(n)}$, and the eq.
(\ref{EQ028}) governs the localization factors $\xi^{\mp}_{n}(y)$. To decouple the extra dimension factors in (\ref{EQ028}) we can define the operators
\begin{eqnarray}\label{EQ030}
\mc{Q}\equiv-e^{A-B}\frac{d}{dy}+\lambda e^{A}f(y)\hspace{1cm}\mbox{and}\hspace{1cm}\mc{Q}^{\dagger}\equiv e^{A-B}\frac{d}{dy}+\lambda e^{A}f(y).
\end{eqnarray}
In this way, equations in (\ref{EQ028}) can be decoupled as
\begin{eqnarray}
\mc{Q}^{\dagger}\mc{Q}\xi^{\pm}_{n}(y)=m^{2}_{n}\xi^{\pm}_{n}(y).\label{EQ031}
\end{eqnarray}
By performing a comparative with {\it non-relativistic supersymmetric quantum mechanics}, we can understand $\mc{H}=\mc{Q}^{\dagger}\mc{Q}$ as a `Hamiltonian' and it is a Hermitian operator. Therefore, the mass values $m^{2}_{n}$ are positive-definite and there are no tachyonic modes in the braneworld model. Beyond this, the solutions $\xi^{\pm}_{n}$ must satisfy the orthonormalization condition
\begin{eqnarray}\label{EQ032}
\int_{-\infty}^{+\infty} dy\xi^{\pm}_{n}\xi^{\pm}_{l}=\delta_{nl}\hspace{1cm}\mbox{and}\hspace{1cm}\int_{-\infty}^{+\infty} dy\xi^{+}_{n}\xi^{-}_{l}=0.
\end{eqnarray}
General approach presented above is that commonly used in the literature.

Beyond this, by using the field configuration (\ref{EQ026}) and the eq. (\ref{EQ028}) , the five dimensional action (\ref{EQ019}) can be factored as
\begin{eqnarray}\label{EQ029}
S^{(m)}&=&-\sum_{n,l}K_{ln}^{+}\int d^{4}x\sqrt{-\hat{g}}\left[i\bar{\psi}^{+}_{(l)}\hat{\Gamma}^{\mu}\hat{D}_{\mu}\psi^{+}_{(n)}+m_{n}\bar{\psi}^{+}_{(l)}\psi^{-}_{(n)}\right]-\nonumber
\\&&-\sum_{n,l}K_{ln}^{-}\int d^{4}x\sqrt{-\hat{g}}\left[ i\bar{\psi}^{-}_{(l)}\hat{\Gamma}^{\mu}\hat{D}_{\mu}\psi^{-}_{(n)}+m_{n}\bar{\psi}^{-}_{(l)}\psi^{+}_{(n)}\right].
\end{eqnarray}
where
\begin{equation}\label{env}
 K_{ln}^{\pm} \equiv \int_{-\infty}^{\infty}e^{B-A}\xi^{\pm}_{l}\xi^{\pm}_{n}dy
\end{equation}

Finally, the confinement can be discussed by looking for solutions of equation (\ref{EQ028}) and performing the integral in extra dimension  (\ref{env}).
As showed in \cite{Freitas:2020mxr}, the finitness of extra dimension integral (\ref{env}) is not sufficient to ensure a consistent confinement of the matter fields. There, by using the Einstein equation, two conditions that the stress tensor of the matter fields must satisfy in order that the localization can be performed consistently. For our case, these conditions can be written as
\begin{eqnarray}\label{EQ038}
T^{m}_{\mu\nu}(x,y)=\frac{\kappa_{4}^{2}}{\kappa_{5}^{2}}\hat{T}^{m}_{\mu\nu}(x),\hspace{1cm}\mbox{and}\hspace{1cm}T^{m}_{55}(x,y)=\frac{1}{2}e^{2(B-A)}\hat{T}^{m}(x),
\end{eqnarray}
where $T^{m}_{\mu\nu}(x,y)$ and $T^{m}_{55}(x,y)$ are the components of the stress tensor of the spinor field obtain from (\ref{EQ019}), and $\hat{T}^{m}(x)=\hat{g}^{\mu\nu}(x)\hat{T}^{m}_{\mu\nu}(x)$. These conditions emerge from the claim that the solutions for $A(y)$ and $B(y)$ should not be changed when we add the spinor field on the bulk and it can be confined. Since the functions $A(y)$ and $B(y)$ are not modified, we can ensure that the gravity (zero-mode) will still remain confined.

\subsection{Confinement of the zero-mode}\label{Sec-3-1}

In this subsection, let us discuss the confinement of the zero-mode, i.e., we will solve equation (\ref{EQ028}) by considering $m_{n}=0$. For this case, the equation (\ref{EQ028}) can be written as
\begin{eqnarray}
-e^{A-B}\frac{d\xi^{\pm}_{0}(y)}{dy}\pm \lambda e^{A}f(y)\xi^{\pm}_{0}(y)=0\label{EQ033}
\end{eqnarray}
and, from equations (\ref{EQ029}) and (\ref{env}), we get the effective action for the zero-mode and it is given by
\begin{eqnarray}\label{EQ034}
S_{0}^{(m)}=-\int d^{4}x\sqrt{-\hat{g}}\left[K_{0}^{+}i\bar{\psi}^{+}_{(0)}\hat{\Gamma}^{\mu}\hat{D}_{\mu}\psi^{+}_{(0)}+K_{0}^{-}i\bar{\psi}^{-}_{(0)}\hat{\Gamma}^{\mu}\hat{D}_{\mu}\psi^{-}_{(0)}\right],
\end{eqnarray}
where
\begin{eqnarray}\label{EQ035}
K^{\pm}_{0}\equiv\int_{-\infty}^{+\infty} dye^{B-A}\left(\xi^{\pm}_{0}\right)^{2}.
\end{eqnarray}
Therefore, effective theory for $\psi^{\pm}_{(0)}$ on the brane will be well-defined when the integrals in extra dimension, (\ref{EQ035}),
is finite. In this way, we said that the zero-mode is localized. This is the {\it finite integral argument} and, to apply it, we need the zero-mode solutions $\xi^{\pm}_{0}$.

Since the eqs. (\ref{EQ033}) are a first order EDO, they can be solved by performing a simple integration in extra dimension. By doing this, we get the zero-mode solutions
 \begin{eqnarray} \label{EQ036}
\xi_{0}^{\pm}(y)&=&c_{\pm}\exp\!\left[\pm\lambda \int_{y} dze^{B(z)}f(z)\right].
\end{eqnarray}
With these solutions, the integrals $K^{\pm}_{0}$ can be analyzed and the confinement can be discussed. For example, for the $5$ dimensional braneworld model presented by M. Gremm in ref. \cite{Gremm:1999pj}, the warp factor is $A(y)=\ln\!\left[\sech^{b}(ay)\right]$ with $b$ a positive parameter and, without the {\it dilaton}, $B(y)=0$. For this model, the integral $K^{\pm}_{0}$ stay
\begin{eqnarray}\label{EQ037}
K^{\pm}_{0}= c^{2}_{\pm}\int_{-\infty}^{+\infty} dy\cosh^{b}(ay)\exp\left[\pm2\lambda \int_{y} dzf(z)\right].
\end{eqnarray}
From this, we see that the free case ($\lambda=0$) cannot be confined, but when $\lambda \neq 0$, we can properly choose the function $f(y)$ to provide the confinement. For most cases in thick brane scenarios, the coupling function, $f(y)$, is the scalar field which generates the brane, $\phi(y)$   \cite{Liu:2008pi, Liu:2009ve,Liu:2009uca,Zhao:2010mk,Castro:2011pp, Cruz:2011ru}. It is worth noting that it is not possible to confine both chiralities, $\psi^{+}_{(0)}$ and $\psi^{-}_{(0)}$, simultaneously. The above results are obtained without the {\it dilaton}, but when this scalar field is present in brane model, $B(y)=rA(y)$ with $r\in[0,1]$ and the results are analogous. For this case, once again, the free field ($\lambda=0$) cannot be localized and, by properly choose the function $f(y)$, the confinement can be achieved for one of the chiralities. Similar results are obtained for other braneworld models \cite{Kehagias:2000au, Koley:2008dh, Castro:2010au, Mendes:2017hmv}. About the confinement of the massive modes, it is often discussed in a superficial way. This because, for most cases, these modes cannot be confined, but other aspects of the theory can be discussed, for example, look for resonant massive modes \cite{Mendes:2017hmv}.

As discussed in the previous section, the finitness of the integral in eq. (\ref{EQ035}) does not ensure that the zero-mode can be confined in a consistent way. Thus, we can apply the consistency conditions, eq. (\ref{EQ038}), for this case. First, we must calculate the stress tensor for the action (\ref{EQ019}),and, we will get
\begin{eqnarray}\label{EQ039}
T^{m}_{MN}(x,y)=i\Bar{\Psi}(x,y)\Gamma_{(M}D_{N)}\Psi(x,y)+g_{MN}\mc{L}^{m}(x,y).
\end{eqnarray}
Now, we can use the solutions obtained in last section and write, for the zero-mode, the components
\begin{eqnarray}\label{EQ040}
T^{m,0,\pm}_{\mu\nu}(x,y)=e^{-3A(y)}(\xi_{0}^{\pm})^{2}\left[i\Bar{\psi}^{\pm}_{0}(x)\hat{\Gamma}_{(\mu}\hat{D}_{\nu)}\psi^{\pm}_{0}(x)+\hat{g}_{\mu\nu}\hat{L}_{0}^{m,\pm}(x)\right],
\end{eqnarray}
where
$$\hat{L}^{m,\pm}(x)=-i\Bar{\psi}^{\pm}_{0}(x)\hat{\Gamma}^{\rho}\hat{D}_{\rho}\psi^{\pm}_{0}(x).$$
We do not need worrying about condition for $T^{m}_{jk}(x,y)$, because it is already satisfied when the condition for $T^{m}_{\mu\nu}(x,y)$ is. Therefore, in order to satisfy the consistency condition $T^{m}_{\mu\nu}(x,y)=\hat{T}^{m}_{\mu\nu}(x)$, we must require that
\begin{eqnarray}\label{EQ041}
e^{-3A(y)}(\xi_{0}^{\pm})^{2}=constant.
\end{eqnarray}
Thus, by using the zero-mode solutions (\ref{EQ036}), the above condition can be written in the equivalent way
\begin{eqnarray}\label{EQ042}
-3A(y)\pm2\lambda \int_{y} dze^{B(z)}f(z)=constant.
\end{eqnarray}
From this relation, we get a strong constraint on the scalar-coupling function $f(y)$. In fact, by consistency reasons, this function must be
\begin{eqnarray}\label{EQ043}
f(y)=f^{+}=\frac{3}{2\lambda}\frac{dA(y)}{dy}e^{-B(y)}\hspace{0.5cm}\mbox{or}\hspace{0.5cm}f(y)=f^{-}=-\frac{3}{2\lambda}\frac{dA(y)}{dy}e^{-B(y)}
\end{eqnarray}
 Just like the confinement is possible only for one of the chiralities, $\psi^{+}_{0}(x)$ or $\psi^{-}_{0}(x)$, the consistency is also  possible only for one of them. We must use $f(y)=f^{+}$ when $\psi^{+}_{0}(x)$ is localized and $f(y)=f^{-}$ when $\psi^{-}_{0}(x)$ is confined. This result is very interesting since it eliminate the arbitrariness in choose the function $f(y)$. As we can see, the consistency conditions imposes a strong constraint in the interactive term $f(y)$, which can be determined by the metric warp factors, i.e., by the brane scenario. In this way, in the next sections, we will test the different brane scenario and compare with the literature results.

\section{Thin brane model in $5$ dimensions}\label{Sec-3}

In previous section we found the scalar-coupling function $f(y)$, eq. (\ref{EQ043}), which localize the zero-mode of the fermion fields, for a generic five dimensional braneworld scenario. To verify how it works in a specific one, in this section we will apply the results for the Randall-Sundrum thin brane model \cite{Randall:1999vf, Randall:1999ee}. This model was built to solve the hierarchy problem, i.e., the gap between the electroweak and the Planck energy scale. To perform the unification of the energy scales, Lisa Randall and Raman Sundrum propose a non-compact extra dimension with a negative cosmological constant, $\Lambda$, described by the action
\begin{equation}
 S = \int \sqrt{-g}d^{4}x\left[\frac{1}{2\kappa_{5}}(R - 2\Lambda) + \mathcal{L}_{b}\right],
\end{equation}
where $\mathcal{L}_{b}$ is the lagrangian density of the brane.
 The model is characterized by the metric
\begin{equation}
 ds^{2} = e^{-2k|y|}\eta_{\mu\nu}dx^{\mu}dx^{\nu} + dy^{2},
\end{equation}
where $k = \sqrt{-\Lambda/6}$ is a constant, and $y$ is the extra dimension. To sustain the above metric, the model require a thin brane localized at $y=0$ with a fine-tuned density and pressure
\begin{eqnarray}
 \rho = -p = -\Lambda/\kappa_{5}.
\end{eqnarray}
When compared to eq. (\ref{EQ018}), we conclude that $A(y) = -k|y|$  and $B(y) = 0$, thus to localize the fermion field in a gravity-consistent way in the Randall-Sundrum brane the result of the subsection \ref{Sec-3-1} constraint that the function $f(y)$ must to be
\begin{eqnarray}\label{RScond}
F^{\pm}(y) \equiv \lambda f^{\pm}= \mp\frac{3k}{2}\sgn(y).
\end{eqnarray}
The authors in Ref. \cite{Barbosa-Cendejas:2015qaa}, approach the  localization of spin $1/2$ on five dimensional brane and got  the Yukawa coupling $F(y) = -Mk\sgn(y)$, where $M$ is their free coupling parameter. In order to obtain a relation between the left chiral ground state and the zero-mode of gravity, the authors choose $M = 3/2$, which agree with our approach and with the result (\ref{RScond}). The above result obtained in this sections must to be recovered for all thin brane model asymptotically AdS in $y \to \infty $ limit.

\section{Thick Brane with a Piecewise Warp Factor}\label{Sec-4}

To soften the divergence produced by thin brane solutions, some models with a piecewise warp factor were proposed \cite{Cvetic:2008gu, Landim:2015paa}. In this section, we will consider the gravity-consistent localization of the fermion field  in brane scenarios with the following warp factor form
\begin{equation}
 A(y) = B(y) =  \left\lbrace \begin{matrix}
                        -\ln(g(y)) & , \mbox{for} \;\;|y| < d\\
                        -\ln\left[k_{0}(|y| +\beta)\right] & , \mbox{for} \;\;|y| > d
                       \end{matrix}
\right.
\end{equation}
where $d$, $\beta$, $k_{0} = \sqrt{-\Lambda/6}$ are positive constants and $g(r)$ a smooth positive and even function in $|y|<d$. The parameter $d$ set the thickness of the brane, and, to be differentiable, the metric, and its first derivative, must to continuous in $y = \pm d$, which restricts the form of the function $g(y)$.  Thus in a generic form we can use the localization condition (\ref{EQ043}) to find the scalar-coupling function in terms of the extra dimension coordinate, $y$, as
\begin{equation}\label{fyPW}
 F^{\pm}(y) =  \mp\left\lbrace \begin{matrix}
                        \frac{3}{2}g'(y) &, \mbox{for}\;\;|y| < d\\
                         \frac{3k_{0}}{2}\;\sgn(y)& ,\mbox{for}\;\; |y| > d
                       \end{matrix}
                       \right.
\end{equation}
As cited before, since the piecewise warp factor brane modifies only the near region, the asymptotic behavior agrees with the Randall-Sundrum one, eq. (\ref{RScond}).

The authors in Ref. \cite{Landim:2015paa} , studies the localization of fermion field using the similar form of above result, but without the factor $3/2$. Since in booth case are no free parameters, their results can not fit (\ref{fyPW}).

Even in piecewise warp factor brane models, the fermion field localization was been studied for \cite{Li:2010dy}
\begin{equation}
 g(y) = \left[\frac{\cos(\sqrt{V_{0}}y) +2}{3}\right]^{-1/2}.
\end{equation}
In order to study the localization of the fermion field, the authors did not consider the coupling function in the form  (\ref{fyPW}), and also they did not consider the gravity-consistency condition, they could not determine the coupling parameter.

\section{Single-field thick brane models}\label{Sec-5}
In this section we will study the gravity-consistent fermion field localization on a thick brane generated by a single real scalar field, $\phi$.  Since a lot of brane models can be produced by this way, we will address it in this section in a general form, and approach the individual models in the following subsections \cite{Liu:2009ve,Guo:2011qt, Liu:2009dw, Melfo:2002wd, Gregory:2001xu, Landim:2015paa, Koley:2004at, Liu:2008pi, Kehagias:2000au,Bazeia:2003aw, Cruz:2011ru, Sorkhi:2018nln, Bazeia:2004dh}. The general setup for these models is given by the action
\begin{equation}\label{actSF}
 S = \int d^{5}x\sqrt{-g}\left[\frac{1}{2\kappa_{5}}\left(R - 2\Lambda\right) -\frac{1}{2}g^{MN}\partial_{M}\phi\partial_{N}\phi -V(\phi)\right],
\end{equation}
where $\kappa_{5} = 8\pi G_{5}$ with $G_{5}$ the five-dimensional Newton constant, and $\Lambda$ is the bulk cosmological constant.
Taking the functional variation of the action (\ref{actSF}) with respect to the metric, $g^{MN}$, and to the scalar field, $\phi$, respectively, we obtain the equations of motion
\begin{eqnarray}
&&R_{MN}-\frac{1}{2}g_{MN}R +\Lambda g_{MN} = \kappa_{5}T_{MN} \label{einSF}
\\&& \frac{1}{\sqrt{-g}}\partial_{M}\left[\sqrt{-g}g^{MN}\partial_{N}\phi\right] - \frac{d V(\phi)}{d\phi} = 0, \label{SFeom}
\end{eqnarray}
where $T_{MN}$ is the energy-momentum tensor of the scalar field.
The most varied scenarios are obtained through the definition of the potential and the presence and sign of the cosmological constant, and the field equations (\ref{einSF}) and (\ref{SFeom}) must to be solved for each one. Since the field equation solutions and the properties of the brane will depend on the specific setup (\ref{actSF}) the gravity-consistent localization of the fermion field will be studied individually below.

\subsection{The de Sitter Thick brane}

Despite the supernova observation, the majority of high dimensional braneworld models consider a flat metric induced in the brane, but to provide a more realistic scenario some de Sitter brane models were proposed, described by the line element
\begin{equation}\label{dsdS}
  ds^{2} = e^{2A}\left(-dt^{2} +e^{2\beta t}d\vec{x}\cdot d\vec{x} +dy^{2}\right),
\end{equation}
where $\beta > 0$ is the Hubble constant on the brane \cite{Liu:2009ve}. To sustain a de Sitter induced metric in this model the scalar field is subject to the potential
\begin{equation}\label{VdS}
 V(\phi) = \frac{1+3\delta}{2\delta}3\beta^{2}\left(\cos\frac{\phi}{\phi_{0}}\right)^{2(1-\delta)},
\end{equation}
where $\phi_{0} = \sqrt{3\delta(1 -\delta)}$ and $0 < \delta < 1$. As showed in literature, using the metric (\ref{dsdS}), this setup provide the following solutions to the scalar field equation of motion and to the warp factor, eqs.(\ref{SFeom}) and (\ref{einSF}) respectively,  \cite{Guerrero:2002ki, Gass:1999gk}
\begin{eqnarray}
 \phi(y) &=& \phi_{0}\arctan\left[\sin\left(\frac{\beta y}{\delta}\right)\right], \label{phidS}
\\ e^{2A} &=& \cosh^{-2\delta}\left(\frac{\beta y}{\delta}\right). \label{AydS}
\end{eqnarray}
As demonstrated in Ref. \cite{Freitas:2020mxr}, the induced cosmological constant on the brane for a single extra dimension is given by
 \begin{equation}\label{icc}
  \alpha = \left[\frac{1}{2}S(y)  +\Lambda  - \kappa^{2}\mathscr{L}^{b}\right]e^{2A},
 \end{equation}
 were, $\mathscr{L}^{b}$
is the lagrangian density of the scalar field which produces the brane and $S(y) = 6[ \nabla_{\mu}\nabla^{\mu}A(y) +2\nabla^{\mu} A(y)\nabla_{\mu}A(y)]$. Using the potential (\ref{VdS}), the scalar field (\ref{phidS}) and the metric (\ref{AydS}) it is easy to show that the model is self consistent, i.e., the induced cosmological constant on the brane, by eq. (\ref{icc}), is given by $\alpha = 3\beta^{2}$, which is consistent with the metric (\ref{dsdS}).

Thus, the gravity-consistent condition, eq.(\ref{EQ043}), provides, for this specific scenario, the scalar-coupling function
\begin{eqnarray}\label{fydS}
{F}^{\pm}(y) =\mp\frac{3}{2}\beta\tanh\left(\frac{\beta y}{\delta}\right) \cosh^{\delta}\left(\frac{\beta y}{\delta}\right)
\end{eqnarray}
% \[\tanh(\phi/\phi_{0}) = \sinh\left(\frac{\beta y}{\delta}\right)\;\;\; \cosh\left(\frac{\beta y}{\delta}\right) = \cos^{-1}(\phi/\phi_{0})\]
or, in terms of the scalar field $\phi$, eq. (\ref{phidS}),
\begin{eqnarray}
\bar{F}^{\pm}(\phi) =\mp\frac{3}{2}\beta\tanh(\phi/\phi_{0})\cos^{1- \delta}(\phi/\phi_{0}). \label{fphi1}
\end{eqnarray}
The localization of the fermion field in this scenario was studied in previous paper, but the authors do not consider the scalar-coupling function (\ref{fphi1}) \cite{Liu:2009ve}.
The de Sitter brane model approached in Ref. \cite{Guo:2011qt}, can be fitted by the limit $\delta \to 1$. The authors examine the fermion field localization in this model and use the specific form (\ref{fydS}), but the authors show that no one chirality can be localized on the brane, due the integral (\ref{EQ036}) diverges.

\subsection{The Symmetric Thick Brane}
Now we will consider a two-parameters family of symmetric thick brane solutions formed by a scalar field. This type of brane is very interesting because it has a very rich internal structure. To perform the model the real scalar field must to be subject to the potential
\begin{equation} \label{VAsy}
  V(\phi) = \frac{3}{2}\lambda^{2}\sin^{2-\frac{2}{s}}\left(\phi/\phi_{0}\right)\cos^{2}\left(\phi/\phi_{0}\right)\left[2s -1 -4\tan^{2}\left(\phi/\phi_{0}\right)\right],
\end{equation}
where $\phi_{0} = \sqrt{3(2s-1)}/s$, $\lambda$ is a positive real constant and $s$ is a positive odd integer \cite{Liu:2009dw}. Under these condition, the brane generated by the above potential are a domain wall, but the parameter $s$ could not be identified with the wall’s inverse thickness. For $s=1$, in five dimensions, the model has been presented in Ref.\cite{Gremm:1999pj} and, by a coordinate transformation, can be charted in a regularized Randall-Sundrum brane.
In this family of flat brane models the five dimensional line element is written in the conformal form
\begin{equation}
 ds^{2} = e^{2A}\left(\eta_{\mu\nu}dx^{\mu}dx^{\nu} +dy^{2}  \right),
\end{equation}
The scalar field equation for the given potential, (\ref{VAsy}), and the Einstein's equation provides as solution \cite{Melfo:2002wd, Gregory:2001xu}
\begin{eqnarray}
\phi &=& \phi_{0}\arctan\left(\lambda y\right)^{s}
 \\e^{2A} &=& \left[1 +(\lambda y)^{2s}\right]^{-1/s}.
\end{eqnarray}
To perform the localization of the fermion field in a gravity-consistent way, the condition  (\ref{EQ043}) constraint the scalar-coupling function to
\begin{eqnarray}
{F}^{\pm}(y) =\mp\frac{3\lambda(\lambda y)^{2s-1}}{2\left[1 +(\lambda y)^{2s}\right]^{(2s-1)/2s}}.
\end{eqnarray}
As commented before, since the model is asymptotic AdS, the above solution have the same behavior of the Randall-Sundrum one.
Writing as a function of the scalar field, we obtain
\begin{eqnarray}\label{fphiSy}
\bar{F}^{\pm}(\phi) =\mp\frac{3}{2}\lambda\sin^{\frac{2s-1}{s}}(\phi/\phi_{0}).
\end{eqnarray}
The localization of fermion field coupled to the scalar field had been studied in the literature, the authors in Ref.\cite{Liu:2009dw} did not consider the specific form (\ref{fphiSy}). For $s=1$ this model could be mapped in smooth warp factor used in Ref. \cite{Landim:2015paa}, where the authors consider the scalar-coupling function (\ref{fphiSy}), fixing their free parameter at $\eta = \mp 3\lambda/2$, where the minus signal localize the right chirality while the plus signal localize the left one. Although the authors did not consider the specific case, $\eta/\lambda = -3/2$, they showed that the right chiral massless fermion is localized if $\eta/\lambda < -1/2$, which agree with the result obtained in this subsection.

\subsection{Sine-Gordon Kink thick brane}
In 2005, R. Koley and S. Kar constructed a sine-Gordon kink class of thick brane  \cite{Koley:2004at}. The kink solution of the scalar field in a warped bulk can provide a thin brane model and to sustain this solution  this model consider a generalized sine-Gordon potential
\begin{equation}
 V(\phi) = p\left( 1 +\cos\frac{2\phi}{q}\right),
\end{equation}
where $p$ and $q$ are free real and positive parameters \cite{Liu:2008pi}. Also the above potential the model consider a non-null 5-dimensional cosmological constant, $\Lambda$, and the following form to the metric
\begin{equation}
 ds^{2} = e^{2A}\eta_{\mu\nu}dx^{\mu}dx^{\nu} +dy^{2}.
\end{equation}
The fields equations provides, for the scalar field and for the warp factor, the solutions
\begin{eqnarray}
\phi &=& 2q\arctan(e^{ky}) - \frac{\pi q}{2} \label{phiSG}
\\ A(y) &=& - \tau \ln \cosh(ky),
\end{eqnarray}
where
$\tau = \frac{1}{3}\kappa_{5}q^{2}$ and $k = \frac{\sqrt{6|\Lambda|}}{6\tau}$.
In this model the parameters $q$ and $\Lambda$ are kept free and $p$ is fixed at
\begin{equation}
 p = \frac{|\Lambda|}{2\kappa_{5}}\left(\frac{\kappa_{5}}{3} +\frac{1}{4q^{2}}\right).
\end{equation}
Now, that we know the warp factor, we can use the condition (\ref{EQ043}) to write the only scalar-coupling function which provides a gravity-consistent localization for the zero-mode of fermion field as
\begin{eqnarray}
{F}^{\pm}(y) =\mp\frac{3k\tau}{2}\tanh(ky).
\end{eqnarray}
As the bulk have a negative cosmological constant and the energy of the scalar field vanishes, the above result returns to the Randall-Sundrum one, eq. (\ref{RScond}), at infinity.

As we know the form of the scalar field, given by eq. (\ref{phiSG}), the above result can be written as a Yukawa coupling function
\begin{eqnarray}
\bar{F}^{\pm}(\phi) =\pm\frac{3k\tau}{2}\cos\left(\frac{2\phi + \pi q}{2q}\right). \label{fphiSG}
\end{eqnarray}
The authors in Ref. \cite{Brihaye:2012um}  studied the localization of fermion field in this scenario, but they only consider the linear coupling case, which can not be by the above result, by a convenient choice of the free parameters. In the paper \cite{Liu:2008pi}, the authors  consider the scalar-coupling function given by $f(\phi) = \sin\phi$, which agree with (\ref{fphiSG}) only in $q =1$ case. In this case, their free coupling parameter $\eta$ could be fixed as $\eta = -\frac{3k\tau}{2}$ and, although they did not consider the gravity-consistent condition, the qualitative result could be recovered, i.e., for a positive $\eta$ only the left chirality can be localized.

The localization of fermion field was studied in an equivalent brane model by Koly and Kar, but they not consider the scalar-coupling function given by (\ref{fphiSG}) \cite{Koley:2004at}.

\subsection{Smooth Brane Generated by a Bounce}\label{5D}

In this section we will consider a brane model generated by a bounce-type configurations of the scalar field \cite{Kehagias:2000au}. The bounce solution are linked to a smooth warp factor, which reduce to the Randall-Sundrum brane in a appropriated limit.  To perform the model the scalar field is subjected to the potential
\begin{equation}
 V(\phi) = \frac{\lambda}{4}\left( \phi^{2} - v^{2}\right)^{2} -\frac{\lambda }{27}k_{5}\phi^{2}\left( \phi^{2} - 3v^{2}\right)^{2},
\end{equation}
where $\lambda$ and $v$ are the free parameters of the model, since it consider a vanish cosmological constant $\Lambda$. The scalar field $\phi(y)$, and the warp factors $A(y)$ and $B(r)$ in eq. (\ref{EQ018}), are given by
\begin{eqnarray}
 \phi(y) &=& v\tanh (ay) \label{phiBonce},
 \\A(y) &=& -\beta\ln\cosh^{2}(ay) - \frac{\beta}{2}\tanh^{2}(ay)
\end{eqnarray}
and $B(r) = 0$, where $a^{2} \equiv \lambda v^{2}/2$. As made in previous sections, the result (\ref{EQ043}) allow us to determine the  scalar-coupling function, which is
\begin{eqnarray}
{F}^{\pm}(y) = \mp\frac{3a\beta}{2}\tanh(ay)\left(2 +\sech^{2}(ay)\right).
\end{eqnarray}
Although the model does not have a cosmological constant the bulk are asymptotically AdS, due the scalar field potential, and the scalar-coupling function agree with the thin brane one at infinity.
In order to produce a coupling between the scalar field and the fermion field, we can white the above result, using the solution (\ref{phiBonce}), as
\begin{eqnarray}\label{fphiBounce}
\bar{F}^{\pm}(\phi)=\mp \frac{9a\beta}{2\nu}\phi\left(1 -\frac{\phi^{2}}{3\nu^{2}}\right).
\end{eqnarray}
Despite the fermion field localization in a smooth brane generate by a bounce has been addressed in the literature, the author consider only the linear scalar-coupling functions \cite{Kehagias:2000au, Mendes:2017hmv}, thus no one parameter can recover the result (\ref{fphiBounce}).

\subsection{Deformed Thick Brane Model}
The next brane model driven by a single scalar field addressed in this study is the deformed kink brane model. The deformation provide an internal structure and enriches the phenomenology of the model.  To provide this scenario, the field are subjected to the potential
\begin{equation}
V(\phi) = V_{p}(\phi) = \frac{1}{8}\left(\frac{d W_{p}}{d\phi}\right) -\frac{1}{3}W_{p}(\phi)^{2},
\end{equation}
where
\begin{equation}\label{Wp}
 W_{p}(\phi) =  \frac{2p}{2p -1}\phi^{\frac{2p-1}{p}} -\frac{2p}{2p +1}\phi^{\frac{2p+1}{p}},
\end{equation}
in a bulk with a vanish cosmological constant. The odd integer parameter $p$ controls the deformation of the kink brane, being the undeformed case restored for $p=1$ \cite{Bazeia:2003aw, Cruz:2011ru, Sorkhi:2018nln, Bazeia:2004dh}. Under the above potential, the solution of the scalar field is given by
\begin{equation}\label{phiyDefor}
 \phi = \tanh^{p}\left(\frac{y}{p}\right),
\end{equation}
and the metric warp factor is given by $B(y) = 0$ and
\begin{equation}\label{AyDefor}
 A(y) =  -\frac{1}{3}\frac{p}{2p +1}\tanh^{p}\left(\frac{y}{p}\right) - \frac{2}{3}\left(\frac{p^{2}}{2p +1} - \frac{p^{2}}{2p -1}\right)\times\left\lbrace \ln\left[\cosh\left(\frac{y}{p}\right)\right] -\sum_{n=1}^{p-1}\frac{1}{2n}\tanh^{2n}\left(\frac{y}{p}\right)\right\rbrace.
\end{equation}
As it was too wide, is more convenient to White the derivative of the warp factor as a function of the scalar field and obtain the scalar-coupling function directly as function of $\phi$. As showed in Ref.\cite{Bazeia:2003aw}, the derivative of warp factor relates with $W_{p}(\phi)$, as
\begin{equation}
 \frac{dA(y)}{dy} = -\frac{1}{3}W_{p}(\phi) =  -\frac{p}{3(2p -1)}\phi^{\frac{2p-1}{p}} +\frac{p}{3(2p +1)}\phi^{\frac{2p+1}{p}}
\end{equation}
where we used the eq. (\ref{Wp}). Now we can use the condition (\ref{EQ043}), to obtain the scalar-coupling function
\begin{eqnarray}\label{fphiDefor}
\bar{F}^{\pm}(\phi) =\mp\left[\frac{p}{2(2p -1)}\phi^{\frac{2p-1}{p}} -\frac{p}{2(2p +1)}\phi^{\frac{2p+1}{p}}\right].
\end{eqnarray}
Searching in the literature for the localization of fermion field in deformed brane scenario the majority paper do not consider the specific form founded in the present work, but, for $p=1$, the authors in Ref. \cite{Sorkhi:2018nln} consider the case $f(y) = \partial^{n}A(y)$. Their results can be charted in  (\ref{fphiDefor}) for $n=1$ and fixing their free coupling parameter at $\eta = \pm 3/2$, where the plus (minus) signal localizes the right (left) chirality. Both results agree, since they obtained that, for $n=1$, the left chirality is localized if $\eta < -1/2$, and we fix it at $\eta = -3/2$ to provide a gravity-consistent localization.

\section{Multiple Field Brane Models}\label{Sec-6}
In this section we will establish the main features of the $5$ dimensional thick brane model formed by multiple scalar fields using the superpotential method. To do this, let us write the action for gravitational field as
\begin{eqnarray}\label{EQ001}
S^{(g)}=\int\left[\frac{1}{2\kappa^{2}}\left(R-2\Lambda\right)-\frac{1}{2}\p_{M}\phi_{i}\p^{M}\phi_{i}-\frac{1}{2}\p_{M}\pi\p^{M}\pi-V(\phi_{i},\pi)\right]\sqrt{-g}d^{4}xdy,
\end{eqnarray}
where $R$ and $\Lambda$ are the Ricci scalar and the cosmological constant on the bulk, respectively. The $3$-brane will be generated by the scalar fields $\phi_{i}$ and $\pi$ for a given $V(\phi_{i},\pi)$, where the subscript $i =1,\cdots, N$ labels the multiple fields, and these fields depend only the extra dimension $y$. From the above action, we can obtain the equations of motion (EOM)
\begin{eqnarray}
G_{MN}+g_{MN}\Lambda =\kappa^{2}\left[\p_{M}\phi_{i}\p_{N}\phi_{i}+\p_{M}\pi\p_{N}\pi+g_{MN}\mc{L}_{b}(\phi_{i},\pi)\right],\label{EQ002}\\
\frac{1}{\sqrt{-g}}\p_{M}\left[\sqrt{-g}\p^{M}\chi\right]=\frac{\p V(\phi_{i},\pi)}{\p \chi},\hspace{0.2cm}\chi=\phi_{i},\pi.\label{EQ003}
\end{eqnarray}
Now, let us consider an ansatz for the warped metric given by
\begin{eqnarray}\label{EQ004}
ds^{2}=g_{MN}dx^{M}dx^{N}=e^{2A(y)}\hat{g}_{\mu\nu}(x)dx^{\mu}dx^{\nu}+e^{2B(y)}dy^{2}
\end{eqnarray}
where, in this metric, the warp factor $A(y)$ and the function $B(y)$ depend only on the extra dimension $y$,and the components $\hat{g}_{\mu\nu}(x)$ are the metric on the brane. With this ansatz, equation (\ref{EQ002}) can be written as
\begin{eqnarray}
\hat{G}_{\mu\nu}(x)+3\hat{g}_{\mu\nu}e^{2A-2B}\left[2A'^{2}+A''-A'B'\right]+\hat{g}_{\mu\nu}\Lambda e^{2A} =\kappa^{2}\hat{g}_{\mu\nu}e^{2A}\mc{L}_{b}(\phi_{i},\pi),\label{EQ005}\\
-\frac{1}{2}\hat{R}(x)e^{2B-2A}+6A'^{2}+e^{2B}\Lambda =\kappa^{2}\left[\phi_{i}'^{2}+\pi'^{2}+e^{2B}\mc{L}_{b}(\phi_{i},\pi)\right].\label{EQ006}
\end{eqnarray}
Now, because the fields $\phi_{i}$ and $\pi$ are functions only the extra dimension $y$, we can perform a separation of the variables in equation (\ref{EQ005}). Namely,
\begin{eqnarray}
\hat{G}_{\mu\nu}(x)+\hat{g}_{\mu\nu}\alpha=0,\label{EQ007}\\
3A''-3A'B'+6A'^{2}+\Lambda e^{2B}-\alpha e^{2B-2A}=\kappa^{2}e^{2B}\mc{L}_{b}(\phi_{i},\pi),\label{EQ008}\\
6A'^{2}+e^{2B}\Lambda-2\alpha e^{2B-2A}=\kappa^{2}\left[\phi_{i}'^{2}+\pi'^{2}+e^{2B}\mc{L}_{b}(\phi_{i},\pi)\right],\label{EQ009}
\end{eqnarray}
where the constant parameter $\alpha$ can be interpreted as an effective cosmological constant, but in this sections we are interested in solutions for the metric where $\alpha=0$. This requirement implies that the metric on the brane will be flat, i.e., $\hat{g}_{\mu\nu}(x)=\eta_{\mu\nu}$. Beyond this, equations (\ref{EQ008}) and (\ref{EQ009}) can be simplified as
\begin{eqnarray}
-3A''+3A'B'=\kappa^{2}\left[\phi_{i}'^{2}+\pi'^{2}\right],\label{EQ010}\\
6A'^{2}+e^{2B}\Lambda=\kappa^{2}\left[\phi_{i}'^{2}+\pi'^{2}+e^{2B}\mc{L}_{b}(\phi_{i},\pi)\right],\label{EQ011}
\end{eqnarray}
and, in addition, the equations of motion for the scalar fields, eq. (\ref{EQ003}), will be written as
\begin{eqnarray}\label{EQ012}
e^{-2B}\left[\chi''+(4A'-B')\chi'\right]=\frac{\p V(\phi_{i},\pi)}{\p \chi},\hspace{0.2cm}\chi=\phi_{i},\pi.
\end{eqnarray}
Here, we can use the superpotential method presented in \cite{Fu:2011pu, Liu:2013kxz}, where, in this approach, the potential is written as
\begin{eqnarray}\label{EQ013}
V(\phi_{i},\pi)=\exp\left(\sqrt{\frac{4r\kappa^{2}}{3}}\pi\right)\left[\frac{1}{2\kappa^{4}}\left(\frac{\p \mc{W}(\phi_{i})}{\p \phi_{i}}\right)^{2}-\frac{(4-r)}{6\kappa^{2}}\mc{W}^{2}(\phi_{i})\right],
\end{eqnarray}
where $r$ a positive parameter. With this definition, the solution can be obtained by
\begin{eqnarray}\label{EQ014}
\frac{dA(y)}{dy}=-\frac{1}{3}\mc{W}(\phi_{i}),\ \frac{d\phi_{i}(y)}{dy}=\frac{1}{\kappa^{2}}\frac{\p \mc{W}}{\p\phi_{i}} \ \ \ \mbox{and}\ \ \ B(y)=-\sqrt{\frac{r\kappa^{2}}{3}}\pi(y)=rA(y).
\end{eqnarray}
Therefore, by defining the superpotential $\mc{W}(\phi_{i})$, we can get the metric solution.

Beyond this, by using relations presented in Eq. (\ref{EQ014}), we can write the scalar-coupling function
\begin{eqnarray}\label{EQ044}
\bar{F}(y)=-\frac{s}{2}\exp\!\left[\sqrt{\frac{r\kappa^{2}}{3}}\pi(y)\right]\!\mc{W}(\phi_{i}),
\end{eqnarray}
and this allows us to write the interaction term in the action (\ref{EQ019}) as
\begin{eqnarray}\label{EQ045}
\mc{L}^{m}_{int}=\frac{s}{2}\exp\!\left[\sqrt{\frac{r\kappa^{2}}{3}}\pi(y)\right]\!\mc{W}(\phi_{i})\Bar{\Psi}\Psi,
\end{eqnarray}
where $s=-1,0$ or $+1$ [$s=-1$ confine $\xi^{-}_{0}$, $s=+1$ confine $\xi^{+}_{0}$ and $s=0$ is the free case]. The superpotential method, allow us to write the Yukawa interaction term in a covariant way. The form (\ref{EQ045}) ensure that a chosen chirality can be localized on a brane in a consistent way with the Einstein equation, without specify the form os the warp factor, as well as the coordinate system.  This is a new result and, in next section, we will analyze the consequences of this for specifics two field brane models.

\subsection{The Asymmetric Two-Field Thick Branes}
To illustrate the result obtained previously, we will evaluate the scalar-coupling function to asymmetric two-field brane model \cite{Zhao:2009ja}. Since are no dilaton field in this model the parameter $r$ must to vanish. The brane are driven by the superpotential
\begin{equation}
 W(\phi,\chi) = 2\phi\left[\lambda\left(\frac{\phi^{2}}{3} - a^{2}\right) +\mu\chi^{2}\right],
\end{equation}
where $\lambda$, $a$ and $\mu$ are the free parameters and we did $\phi_{1} = \phi$ and $\phi_{2} = \chi$. The superpotential method allow us to obtain the scalar-coupling function by (\ref{EQ044})

\begin{eqnarray}
\bar{F}(y) = s\phi\left[\lambda\left(\frac{\phi^{2}}{3} -a^{2}\right) +\mu\chi^{2}\right].
\end{eqnarray}
As showed previously, this is the only coupling which localizes the fermion field on the brane in a gravity-consistent way. Despite the localization of $1/2$ spinorial field on asymmetric two-field brane model has been studied in literature, the authors did not consider the result obtained in the present work \cite{Zhao:2009ja}.

\subsection{The Non-interacting Two-Fields Brane Model }
In 2015, Dutra and co-workers, proposed a thick brane model driven by multiple scalar fields \cite{deSouzaDutra:2014ddw}. The non-interacting two-field brane model could be understood as a generalization of kink brane for two fields \cite{Farokhtabar:2016fhm}. The superpotential which drives the model can be written as
\begin{equation}
 W(\phi_{1},\phi_{2}) = \lambda_{1}\left(\phi_{1} - \frac{\phi_{1}^{3}}{3}\right) +\lambda_{2}\left(\phi_{2} - \frac{\phi_{2}^{3}}{3}\right)
\end{equation}
where $\lambda_{1}$ and $\lambda_{2}$ are the free parameters. Thus, the main result (\ref{EQ044}) allow us to find the scalar-coupling function
\begin{eqnarray}\label{fphiTF}
\bar{F}(y) = \frac{s}{2}\left[\lambda_{1}\left(\phi_{1} - \frac{\phi_{1}^{3}}{3}\right) +\lambda_{2}\left(\phi_{2} - \frac{\phi_{2}^{3}}{3}\right)\right].
\end{eqnarray}
The authors in Ref. \cite{Farokhtabar:2016fhm} consider some scalar-coupling function in order to localize the fermion field, like $f = \phi_{1}\phi_{2}$ and $f = \phi_{1} +\beta\phi_{2}$. They obtain the conditions which localize the left-hand or the right-hand fermions on the brane, but no one parameter fixation can recover the result   (\ref{fphiTF}).

\subsection{Domain Wall with dilaton coupling}

In this subsection we will study the localization of the fermion field in a smooth brane with a dilaton coupling. The brane model can be mapped in the model studied in subsection \ref{5D}, and can rise form the following superpotential
\begin{equation}
 W(\phi) = ab\phi\left(1 -\frac{\phi^{2}}{3a^{2}}\right),
\end{equation}
where $a$ and $b$ are the free parameters of the model. As we show previously, the single scalar-coupling function, which localizes the fermion field on the brane in a gravity-consistent way, is given by eq. (\ref{EQ044}). For the above superpotential and considering the dilaton coupling, this result leads to

\begin{equation}\label{F5C}
\bar{F}(y)=-s\frac{ab}{2}\phi\left(1 -\frac{\phi^{2}}{3a^{2}}\right)\exp\!\left[\sqrt{\frac{r\kappa^{2}}{3}}\pi(y)\right].
\end{equation}

The authors in Ref. \cite{Mendes:2017hmv} study the localization of the fermion field in this brane scenario, but they only use the scalar-coupling function in the form $F(\phi,\pi) = \eta_{2}\phi\;e^{-\lambda\pi}$. They study the case when $\lambda = \sqrt{r\kappa^{2}/3}$, but they do not consider the specific form (\ref{F5C}). In this sense, no parameter constraint can recover our result and we can not compare our result with their localization method.

\subsection{Block Brane}
The last two-field brane model that we will address in this work is the block brane model \cite{Castro:2010au, Almeida:2009jc,Bazeia:2004dh}. This models also does not have the dilaton field, so that the parameter $r$ must to vanish. In the superpotential formalism we can write
\begin{equation}
 W(\phi,\chi) = 2\phi - \frac{2}{3}\phi^{3} -2a\phi\chi^{2},
\end{equation}
where $a$ is a free parameter and $\phi$ and $\chi$ are the scalar fields. As made in previous cases, the main result given by eq. (\ref{EQ044}) allow us to find the only scalar-coupling function which localizes the fermion field in a gravity-consistent way, which is
\begin{eqnarray}
\bar{F}(y) = s\left(\phi - \frac{1}{3}\phi^{3} -a\phi\chi^{2} \right).
\end{eqnarray}
As the authors of the articles found in the literature did not have the guide given by the equation (\ref{EQ044}), they often use a linear combination of the fields, to study the localization of fermion field \cite{ Almeida:2009jc, Bazeia:2004dh}. In this way they do not consider the form obtained in the present work.

\section{Confinement of massive mode of the spinorial field}\label{Sec-7}

We saw in the subsection (\ref{Sec-3-1}), that the confinement of the spinor field (zero-mode) can be obtained by properly choose the scalar-coupling function $f(y)$. Unfortunately, only the finite integral argument does not remove the ambiguity in defining this function. However, as discussed in last section, some consistency requirements (obtained from Einstein equation) impose a strong constraint on the shape of this function. Below, let us discuss the consequences of this on the massive modes localization.

To discuss the confinement of the massive modes, let us start from equation (\ref{EQ031}), namely,
\begin{eqnarray}\label{EQ046}
e^{2(A-B)}\left[-\frac{d^{2}}{dy^{2}}-(A'-B')\frac{d}{dy}\pm s\frac{3}{2}(A'-B')A'\pm s\frac{3}{2}A''+s^{2}\frac{9}{4}A'^{2}\right]\xi^{\pm}_{n}(y)=m^{2}_{n}\xi^{\pm}_{n}(y).
\end{eqnarray}
Since scalar-coupling $f(y)$ is effectively a function of $A(y)$ and $B(y)$, we will keep the parameter $s$ to explain the contribution of the interaction. Now, let us consider $B(y)=rA(y)$ with $r\in[0,1]$ a numerical parameter \cite{Mendes:2017hmv}. Thus, the above equation stay
\begin{eqnarray}\label{EQ047}
e^{2(1-r)A}\left[-\frac{d^{2}}{dy^{2}}-(1-r)A'\frac{d}{dy}\pm s\frac{3}{2}(1-r)A'^{2}\pm s\frac{3}{2}A''+s^{2}\frac{9}{4}A'^{2}\right]\xi^{\pm}_{n}(y)=m^{2}_{n}\xi^{\pm}_{n}(y).
\end{eqnarray}
As discussed in \cite{Mendes:2017hmv}, we mentioned previously that for $r\in[0,1)$ the energy-momentum tensor ($T_{00}$) of the spinorial field goes to zero in the limit $y\to \infty$. On the other hand, this component goes to a constant negative value for $r=1$. The above equation can be written as a Schrödinger-like equation by proposing
\begin{eqnarray}\label{EQ048}
\xi^{\pm}_{n}(y)=\exp\left[-\frac{(1-r)}{2}A(y)\right]h^{\pm}_{n}(y),
\end{eqnarray}
therefore
\begin{eqnarray}\label{EQ049}
-\frac{d^{2}h^{\pm}_{n}}{dy^{2}}+\left[N_{\pm}(r,s)A''+N_{\pm}^{2}(r,s)\!A'^{2}-m^{2}_{n}e^{-2(1-r)A}\right]h^{\pm}_{n}=0,
\end{eqnarray}
where $2N_{\pm}(r,s)=1-r\pm 3s$. By observing the above equation, we can understand $h^{\pm}_{n}(y)$ as a ``zero-mode'' and an effective ``potential'' can be written as follows
\begin{eqnarray}\label{EQ050}
U^{\pm}_{r,s}(y,m_{n})=N_{\pm}(r,s)A''+N_{\pm}^{2}(r,s)\!A'^{2}-m^{2}_{n}e^{-2(1-r)A}.
\end{eqnarray}
To perform a qualitative discussion about the confinement of the massive modes, let us use the thick brane model presented in reference \cite{Gremm:1999pj}. For this case, the warp factor $A(y)$ is given by
\begin{eqnarray}\label{EQ051}
A(y)=\ln\!\left[\sech^{b}(ay)\right],
\end{eqnarray}
and the above potential stay
\begin{eqnarray}\label{EQ052}
U^{\pm}_{r,s}(y,m_{n})=-ba^{2}N_{\pm}(r,s) \sech^{2}(ay)+b^{2}a^{2}N_{\pm}^{2}(r,s)\tanh^{2}(ay)-m^{2}_{n}\cosh^{2(1-r)b}(ay).
\end{eqnarray}
Remember that $r\in[0,1]$, $b$ is a positive parameter and $m_{n}^{2}\geq0$. Beyond this, there is a symmetry on this potential given by $U^{\pm}_{r,+1}(y,m_{n})=U^{\mp}_{r,-1}(y,m_{n})$. Therefore, by performing the discussion for $s=+1$, the conclusions will be similar to $s=-1$.

The above potential, in the limit $y\to\pm\infty$, goes to
\begin{eqnarray}\label{EQ053}
U^{\pm}_{r,s}(y\to\pm\infty,m_{n})=b^{2}a^{2}N_{\pm}^{2}(r,s)-m^{2}_{n}e^{2(1-r)ba|y|}.
\end{eqnarray}
From this, it is easy to see that for $r\in[0,1)$, this potential diverges for $U^{\pm}_{r,s}(y\to\pm\infty,m_{n})\to-\infty$ and, therefore, this potential cannot provide a stable confinement on the brane. On the other hand, for the particular value $r=1$, the above potential goes to a constant value, namely,
$$U^{\pm}_{1,s}(y\to\pm\infty,m_{n})=b^{2}a^{2}N_{\pm}^{2}(1,s)-m^{2}_{n}$$
For this particular case, asymptotic solutions for $h^{\pm}_{n}$ can be obtained and they are given by
\begin{eqnarray}\label{EQ054}
h_{n}(y_{\infty})\propto \exp\left[\pm\sqrt{b^{2}a^{2}N_{\pm}^{2}(1,s)-m^{2}_{n}}\ y\right].
\end{eqnarray}
Thus, for $m^{2}_{n}<b^{2}a^{2}N_{\pm}^{2}(1,s)$ must exist stable confined massive modes. Beyond this, these confined modes should be discretized. On the other hand, for $m^{2}_{n}>b^{2}a^{2}N_{\pm}^{2}(1,s)$, the above asymptotic solutions become oscillating and, for this continuous modes, the confinement cannot be achieved. Note that for free case ($s=0$) and $r=1$, the coefficient $N_{\pm}^{2}(1,0)=0$, the solutions (\ref{EQ054}) will be oscillating and, therefore, confinement cannot be obtained. Below, let us perform the quantitative discussion for the case with $r=1$.

\begin{minipage}{7.5cm}
\begin{figure}[H]
\center
\includegraphics[scale=0.4]{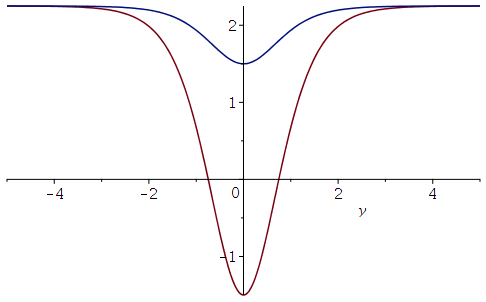}
\caption{Plot of the potentials $U_{+}$ [red] and $U_{-}$ [blue] for $s=+1$ and $b=1$.}\label{Fig01}
\end{figure}
\end{minipage}\hspace{1cm}
\begin{minipage}{7.5cm}
\begin{figure}[H]
\center
\includegraphics[scale=0.4]{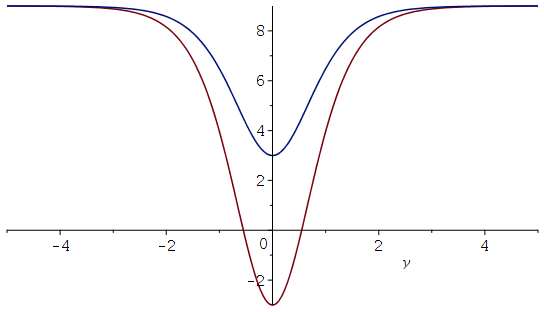}
\caption{Plot of the potentials $U_{+}$ [red] and $U_{-}$ [blue] for $s=+1$ and $b=2$.}\label{Fig02}
\end{figure}
\end{minipage}

\subsection{Massive modes for $r=1$}
As we showed above, the potential for $r=1$ allows us obtain localized massive modes. Now, let us perform the quantitative discussion about this confinement. To do this, let us write below the equation of motion (\ref{EQ049}) with the potential (\ref{EQ052}), i.e.,
\begin{eqnarray}
-\frac{d^{2}h^{+}_{n}}{dy^{2}}+\left[-\frac{3sba^{2}}{2}\sech^{2}(ay)+\frac{9s^{2}a^{2}b^{2}}{4}\tanh^{2}(ay)\right]h^{+}_{n}=-\frac{d^{2}h^{+}_{n}}{dy^{2}}+U_{+}h^{+}_{n}=m^{2}_{n}h^{+}_{n},\label{EQ055}\\
-\frac{d^{2}h^{-}_{n}}{dy^{2}}+\left[\frac{3sba^{2}}{2}\sech^{2}(ay)+\frac{9s^{2}a^{2}b^{2}}{4}\tanh^{2}(ay)\right]h^{-}_{n}=-\frac{d^{2}h^{-}_{n}}{dy^{2}}+U_{-}h^{-}_{n}=m^{2}_{n}h^{-}_{n}.\label{EQ056}
\end{eqnarray}
Figure (\ref{Fig01}) shows a plot of the potential $U_{\pm}$ for $s=+1$. As we can see, this plot present a range of mass, between the minimum and the maximum of $U_{-}$ [blue plot], which must have both chiralities normalizable. Therefore, it must be possible found both chiralities of massive modes confined on the brane. More specifically, for values of mass in the range
$$m_{n}^{2}\in \left[\frac{3ba^{2}}{2},\frac{9b^{2}a^{2}}{4}\right),$$
the localization can be achieved. Furthermore, extrapolating the analogy with the Schrödinger quantum mechanics, it must be possible obtain discretized modes for this system.

The solutions for (\ref{EQ055}) and (\ref{EQ056}) can be obtained by proposing the transformation $u=\tanh(ay)$. By doing this and using $s=+1$, we get
\begin{eqnarray}
\left(1-u^{2}\right)\frac{d^{2}h^{+}_{n}}{du^{2}}-2u\frac{dh^{+}_{n}}{du}+\left[\frac{3b}{2}\left(\frac{3b}{2}+1\right)-\frac{9a^{2}b^{2}-4m^{2}}{4a^{2}\left(1-u^{2}\right)}\right]h^{+}_{n}=0,\label{EQ057}\\
\left(1-u^{2}\right)\frac{d^{2}h^{-}_{n}}{du^{2}}-2u\frac{dh^{-}_{n}}{du}+\left[\frac{3b}{2}\left(\frac{3b}{2}-1\right)-\frac{9a^{2}b^{2}-4m^{2}}{4a^{2}\left(1-u^{2}\right)}\right]h^{-}_{n}=0.\label{EQ058}
\end{eqnarray}
Above equations are associated Legendre equations, namely,
\begin{eqnarray}
\left(1-z^{2}\right)\frac{d^{2}P^{l}_{n}}{dz^{2}}-2z\frac{dP^{l}_{n}}{dz}+\left[n\left(n+1\right)-\frac{l^{2}}{\left(1-z^{2}\right)}\right]P^{l}_{n}=0,\label{EQ059}
\end{eqnarray}
where the parameters $n$ and $l$ must be integers with $0\leq l\leq n$.

Therefore, we get the solutions
\begin{eqnarray}
h_{n,l}^{+}(y)\!\!&=&\!\!c^{+}_{n,l} \mbox{P}_{n}^{l}\!\left[\tanh(ay)\right],\label{EQ060}\\
h_{n,l}^{-}(y)\!\!&=&\!\!c^{-}_{n,l} \mbox{P}_{n-1}^{l}\!\left[\tanh(ay)\right],\label{EQ061}
\end{eqnarray}
with
$$n=\frac{3b}{2}\hspace{0.5cm}\mbox{and}\hspace{0.5cm}l=\frac{\sqrt{9a^{2}b^{2}-4m_{n}^{2}}}{2a}.$$
Now, let us discuss the boundary conditions. By looking at the `potential', we see that the minimum around the brane is a {\it global minimum}. Therefore, there is not possible a particle tunneling out to the bulk. In other words, these localized particles on the brane will still confined indefinitely. Thus, to achieve this, the above confined solutions must be zero on the limit $y\to\pm\infty$. This condition implies that
\begin{eqnarray}
h_{n,l}^{+}(y\to\pm\infty)\!\!&=&\!\!c^{+}_{n,l} \mbox{P}_{n}^{l}\!\left(\pm 1\right)=0,\label{EQ062}\\
h_{n,l}^{-}(y\to\pm\infty)\!\!&=&\!\!c^{-}_{n,l} \mbox{P}_{n-1}^{l}\!\left(\pm 1\right)=0.\label{EQ063}
\end{eqnarray}
These conditions will be satisfied, simultaneously, when $n\geq2$ with $l\neq0$ and $1\leq l\leq (n-1)$. With this, the allowed (confined) values of the mass are given by
\begin{eqnarray}\label{EQ064}
m_{n,l}^{2}=a^{2}\left[n^{2}-l^{2}\right].
\end{eqnarray}
In this way, by using these solutions in the action (\ref{EQ029}), the integral in extra dimension $y$ can be performed and the effective theory on the brane gets
\begin{eqnarray}\label{EQ065}
S^{(spinor)}=-i\int d^{4}x\sqrt{-\hat{g}}\bar{\psi}^{+}_{(0)}\hat{\Gamma}^{\mu}\hat{D}_{\mu}\psi^{+}_{(0)}-\sum_{n}\sum_{l=1}^{n-1}\int d^{4}x\sqrt{-\hat{g}}\left[i\bar{\psi}^{+}_{(n,l)}\hat{\Gamma}^{\mu}\hat{D}_{\mu}\psi^{+}_{(n,l)}+m_{(n,l)}\bar{\psi}^{+}_{(n,l)}\psi^{-}_{(n,l)}\right.\nonumber\\ \left.+i\bar{\psi}^{-}_{(n,l)}\hat{\Gamma}^{\mu}\hat{D}_{\mu}\psi^{-}_{(n,l)}+m_{(n,l)}\bar{\psi}^{-}_{(n,l)}\psi^{+}_{(n,l)}\right]+S^{(spinor)}_{cont.\ modes}.
\end{eqnarray}
Note that $n$ is fixed by the parameter $b$ which is related to the braneworld model. Beyond this, discretized mass values (\ref{EQ064}) depend on the parameter related to the brane thickness, namely, the parameter $a$.

\section{Conclusions}\label{Sec-8}

In this manuscript, we studied the localization of the spin $1/2$ spinor field on a brane. We started by discussing the confinement of the zero-mode. In doing this, we propose the commonly used localization mechanism for thick brane where $\mc{L}_{int}=-\lambda f(y)\bar{\Psi}\Psi$. We showed that the confinement can be achieved by properly chose the function $f(y)$, a result already found in the literature \cite{ Kehagias:2000au, Mendes:2017hmv}. Next, we tested the consistency of the localization procedure studied in reference \cite{Freitas:2020mxr} for a lot of single field brane models and compare the results with the literature. As previous works did not consider the consistency condition, the vast majority of the cases studied the coupling function was not compatible with the obtained form in the present manuscript. In compatible cases, the consistency condition sets the free parameters. In a brane constructed by multiple scalar fields using the superpotential method, by applying the consistency conditions, we found that the function $f(y)$ must have a specific shape given by
$$f(y)=-\frac{s}{2\lambda}\exp\!\left[\sqrt{\frac{r\kappa^{2}}{3}}\pi(y)\right]\!\mc{W}(\phi).$$
With this shape, the confinement of the zero-mode does not change the vacuum solution of the metric and this ensure that the gravity (zero-mode) will still remain localized. Up to now, despite the various coupling functions found in the literature, the above function was not used to study the confinement of the spinor field. Due to this, we used it to discuss the massive modes localization.

By using the above function in equation of motion for the massive modes, we obtained equations (\ref{EQ029}). From this, we found a Schrödinger-like equation and, by analyzing the effective potential, we discussed qualitatively the possibility of the massive modes localization for different values of the parameter $r\in[0,1]$. With this, we showed that massive modes confinement is possible for $r=1$ which implies in $B(y)=A(y)$. For this particular case, we found a range of discretized masses that can be confined on the brane (both chiralities) model presented in \cite{Gremm:1999pj}. As discussed in the previous section, these masses are given by
$$m_{n,l}^{2}=a^{2}\left[n^{2}-l^{2}\right].$$
In the literature, we can find some proposals where the ``confinement'' of the massive modes is achieved, however, for most of then, this confinement is not stable \cite{Melfo:2006hh, Castro:2010uj}. In other words, the confinement is possible, but this modes can tunneling to the bulk and the particle disappears from the brane. For our case, this localization is stable, i.e., the confined masses do not disappears from the brane. In reference \cite{Ringeval:2001cq}, the ``lifetime'' can be made indefinitely large, but for this the fermion must be strongly coupled to the scalar field $f(y)$. In reference \cite{Liu:2013kxz}, the authors indicates qualitatively the massive modes can be confined with a scalar-coupling function given by $f(y)=-\eta e^{\tilde{\lambda}\pi}\phi$. A similar result were obtained in reference \cite{Landim:2015paa}, where the authors define, {\it ad hoc}, a particular function $f(y)$. In this context, our scalar-coupling function emerges in a natural way from consistency requirement with Einstein equation. Beyond this, we found analytical solutions for the discretized modes and exact values for the discretized masses which is closely related to the brane thickness. We believe that, with our results, should be possible to establish constraints on the brane parameter, namely, $a$ and $b$, by using the properties of the known fermions in $4$D. \\

\end{document}